\documentclass{CUP-JNL-EDS}

\usepackage{latexsym}
\usepackage{graphicx}
\usepackage{multicol,multirow}
\usepackage{amsmath,amssymb,amsfonts}
\usepackage{mathrsfs}
\usepackage{amsthm}
\usepackage{rotating}
\usepackage{appendix}
\usepackage[authoryear]{natbib}
\usepackage{ifpdf}
\usepackage[T1]{fontenc}
\usepackage{times}
\usepackage{newtxmath}
\usepackage{textcomp}%
\usepackage{xcolor}%
\usepackage{hyperref}
\usepackage{lipsum}
\jname{Environmental Data Science}
\jyear{2020}
\jvol{4}
\jdoi{10.1017/eds.2020.xx}
\usepackage{wrapfig, caption, subcaption}

\begin{document}

\setlength{\abovedisplayskip}{4pt}
\setlength{\belowdisplayskip}{4pt}

\begin{Frontmatter}
\title[Article Title]{Stochastic Parameterization of Column Physics
  using\\ Generative Adversarial Networks}

\author*[1]{Balasubramanya T. Nadiga}\email{balu@lanl.gov}%\orcid{0000-0002-8514-4315}
\author[1]{Xiaoming Sun}
\author[2]{Cody Nash}

\address[1]{\orgname{Los Alamos National Laboratory},
  \orgaddress{\city{Los Alamos}, \postcode{87545}, \state{NM},  \country{USA}}}

\address*[2]{\orgname{Independent Researcher},
  \orgaddress{\city{Topol'\u cany}, \postcode{95501},  \country{Slovakia}}}

\received{28 January 2022}
%\revised{}
%\accepted{}

\authormark{Nadiga et al.}


\abstract{
 We demonstrate the use of a probabilistic machine learning technique to develop stochastic parameterizations of atmospheric column-physics. After suitable preprocessing of NASA’s Modern-Era Retrospective analysis for Research and Applications, version 2 (MERRA2) data to minimize the effects of high-frequency, high-wavenumber component of MERRA2 estimate of vertical velocity, we use generative adversarial networks to learn the probability distribution of vertical profiles of diabatic sources conditioned on vertical profiles of temperature and humidity. This may be viewed as an improvement over previous similar but deterministic approaches that seek to alleviate both, shortcomings of human-designed physics parameterizations, and the computational demand of the “physics” step in climate models.
}

\end{Frontmatter}

\section*{Impact Statement}
Global climate models can now be used to produce realistic simulations
of climate. However, large uncertainties remain: e.g., the estimated
change in globally-averaged-surface-temperature to a doubling of
atmospheric CO$_2$ varies between $\sim$2 and $\sim$6
degrees-Centigrade across leading models. Uncertainty in representing
cumulus convection (think thunderstorm) is a major contributor to this
spread: Scales at which they occur, 100m-10km, are too small to be
resolved in global climate models, requiring their effects on larger
scales to be approximated with simple models. Improving such
approximations using new probabilistic machine learning techniques,
initial steps towards which are successfully demonstrated in this
work, will likely lead to improvements in the modeling of climate
through improvements in the representation of cumulus convection in
climate models.

\section{Introduction and Problem Formulation}
Over the past 70 years, there has been a concerted effort to develop
first principles based models of Earth’s climate. In this ongoing
effort, current (state-of-the-art, SOTA) comprehensive climate models
and Earth System Models (ESMs) have achieved a level of sophistication
and realism that they are proving invaluable in improving our
understanding of various processes underlying the climate system and
its variability \citep[][pg. 82]{chen2021framing}.  Limited by
computational resources, however, most comprehensive climate
simulations are currently conducted with horizontal grid spacings of
between 50 and 100 km in the atmosphere. However, many atmospheric
processes occur at much smaller scales. These include cumulus
convection, boundary layer turbulence, cloud microphysics, and
others. Even as resolutions continue to slowly increase, a number of
these small-scale processes are not expected to be explicitly resolved
in long-term simulations of climate for a long time. As such, the
effects of such unresolved subgrid processes on the resolved scales
have to be represented using parameterizations.

To wit, the evolution of temperature $T$ and moisture (represented by
specific humidity) $q$ may be written as \citep[e.g.,
see][]{yanai1973determination, nitta1977response}
\begin{align}
\frac{\partial T}{\partial t} + u \frac{\partial T}{\partial x} + v \frac{\partial T}{\partial y} -\omega S_p &= q_1,\\ 
\frac{\partial q}{\partial t} + u \frac{\partial q}{\partial x} + v
  \frac{\partial q}{\partial y} -\omega \frac{\partial q}{\partial p}
  &= - q_2.
\end{align}  
Here, ($u, v$) is the horizontal velocity field, $\omega$ is the
vertical velocity in pressure coordinates, $S_p$ is static stability
given by $S_p = -\frac{T}{\theta}\frac{\partial \theta}{\partial p}$,
where $\theta$ is potential temperature and other notation is standard.  To
simplify notation, in the above equations all variables on the
left hand side (LHS) are understood to be variables at the resolved
scale. With that convention, $q_1$ and $q_2$ are parameterizations
that represent the effect of unresolved subgrid processes on the
evolution of resolved scale temperature and moisture. (Additional mass
and momentum conservation equations complete the system, e.g., by
providing the evolution of ($u, v, \omega$).) For
convenience, following \citep{yanai1973determination}, we refer to
$q_1$ as apparent heating and $q_2$ as apparent moisture sink.

In the above equations, terms on the LHS represent slow, large-scale
dynamical evolution. Phenomenologically, when the large-scale
dynamical evolution or forcing serves to set-up conditions that are
convectively unstable, the system responds by locally and
intermittently forming cumulus convective updrafts that serve to
locally resolve or eliminate the instability.  Indeed, the interaction
between the large-scale dynamics and small-scale intermittent moist
nonhydrostatic cumulus response continues over a range of subgrid
scales, from minutes and hundreds of meters to hours and up to ten
kilometers. Thus, from the point of view of the equations above,
$q_1$ and $q_2$ represent the effect of unresolved cumulus and other
subgrid processes on the evolution of the vertical profiles of
environmental averages of the thermodynamic variables $T$ and $q$;
$q_1$ is an apparent heating source and $q_2$ is an apparent moisture
sink. In particular, $q_1$ comprises effects of radiative and latent
heating and vertical turbulent heat flux, and $q_2$ effects of
latent heating. The effects of horizontal turbulent transport are
smaller in comparison and are therefore commonly neglected, as we do here.

% However, given that  myriad processes occuring on a vast range
% of spatial and temporal scales interact to produce climate,
% computational limitations prevent the first-principles representation
% of numerous such processes in global models of climate. For example,
% convective updrafts that lead to the formation of cumulus clouds begin
% to be resolved as a natural nonhydrostatic feature of stratified flows
% at a horizontal resolution of about 4 km whereas SOTA climate
% simulations have a resolution of between 50 and 100 km. As such
% convective updrafts have to be parameterized as a “subgrid” process in
% such models.

In SOTA climate models, forward evolution of the atmospheric state
over each timestep consists of a “dynamics'' update (solving the mass
and momentum equations and computing terms on the LHS of Eqs.~1 and 2)
followed by a “physics” update. It is the “physics”
update---effectively computing $q_1$ and $q_2$---that implements the
various schemes that capture the effect of unresolved dynamical and
thermodynamic processes (e.g., turbulence transport including that of
convective updrafts, various boundary layer processes, cloud
microphysics, radiation. etc.) on the resolved scales. Physics
parameterizations in SOTA climate models tends to be one of the most
computationally intensive parts \cite[e.g.,
see][]{bosler2019conservative, bradley2019communication}. Furthermore,
since traditional parameterizations are based on our limited
understanding of the complex subgrid-scale processes, significant
inaccuracies persist in the representation of microphysics, cumulus,
boundary layer, and other processes
\citep[e.g.,][]{sun2014high}. Indeed, as pointed out by
\cite{sherwood2014spread}, uncertainty in cumulus parameterization
is a major source of spread in climate sensitivity of
ESMs that are used for climate projections for the 21st century.

The recent explosion of research in machine learning (ML) has
naturally led to the examination of the use of ML in alleviating both,
shortcomings of human-designed physics parameterizations, and the
computational demand of the “physics” step in climate models. In this
context, recent research has focussed on using short high resolution
simulations of climate that resolve convective updrafts, commonly
called cloud resolving models (CRM) as ground truth for developing
a computational-cheaper, unified machine learned parameterization that
replaces the traditional “physics” step. For example, using this
approach, \cite{brenowitz2018prognostic} (BB18) were able to improve the
representation of temperature and humidity fields when compared with
traditional physics parameterizations.

In this study, instead of using output from high-resolution CRMs,
since CRMs can themselves deviate from reality in many aspects
\citep[e.g.,][]{lean2008characteristics, weisman2008experiences,
  herman2016extreme, zhang2016diurnal}, we train a ML model using
reanalysis fields from the Modern-Era Retrospective Analysis for
Research and Applications version 2 (MERRA-2,
\cite{rienecker2011merra}). Further, two similar input profiles (of
temperature and humidity) can evolve differently due to chaotic
subgrid cumulus dynamics and result in different output profiles (of
apparent heating and apparent moisture sink). As such, we estimate
$q_1$ and $q_2$ using a probabilistic ML model. This choice is also
consistent with the suggestion that a stochastic approach is likely
more appropriate for parameterizations in terms of reducing the
uncertainties stemming from the lack of scale separation between
resolved and unresolved processes
\citep[e.g., see][]{nadiga2007instability, duan2007stochastic,
  nadiga2008orientation, palmer2019stochastic}.

\section{The Tropical Eastern Pacific, Data and Preprocessing}
The inter-tropical convergence zone (ITCZ) is one of the
easily-recognized large-scale features of earth's atmospheric
circulation. It appears as a zonally-elongated band of clouds that at
times even encircles the globe. It comprises of thunderstorm systems
with low level convergence, and intense convection and precipitation,
The domain we choose---the region of Eastern Pacific, 15S-15N and
180-100.25W---includes both a segment of the more coherent northern
branch of the ITCZ as a band across the northern part of the domain,
and a segment of the more seasonal southern branch in the southwestern
part of the domain. In the top-left panel of Fig.~1, which show the
2003 annual mean of column-averaged specific humidity, the ITCZ is
contained in the broad and diffuse bands of high moisture seen in
yellow. While not evident in the equivalently averaged temperature
field (top-right), it is better identified in the $q_1$ and $q_2$
fields (bottom row; see later). To aid in geo-locating the domain
considered and the ITCZ, the inset in the bottom-right panel shows the
$q_1$ field for January 1$^{st}$, 2003 on a map with the coastlines of
the Americas.
\begin{figure}[t]%
  \begin{center}
  \includegraphics[trim=0 30 0 20, clip,width=0.8\textwidth]{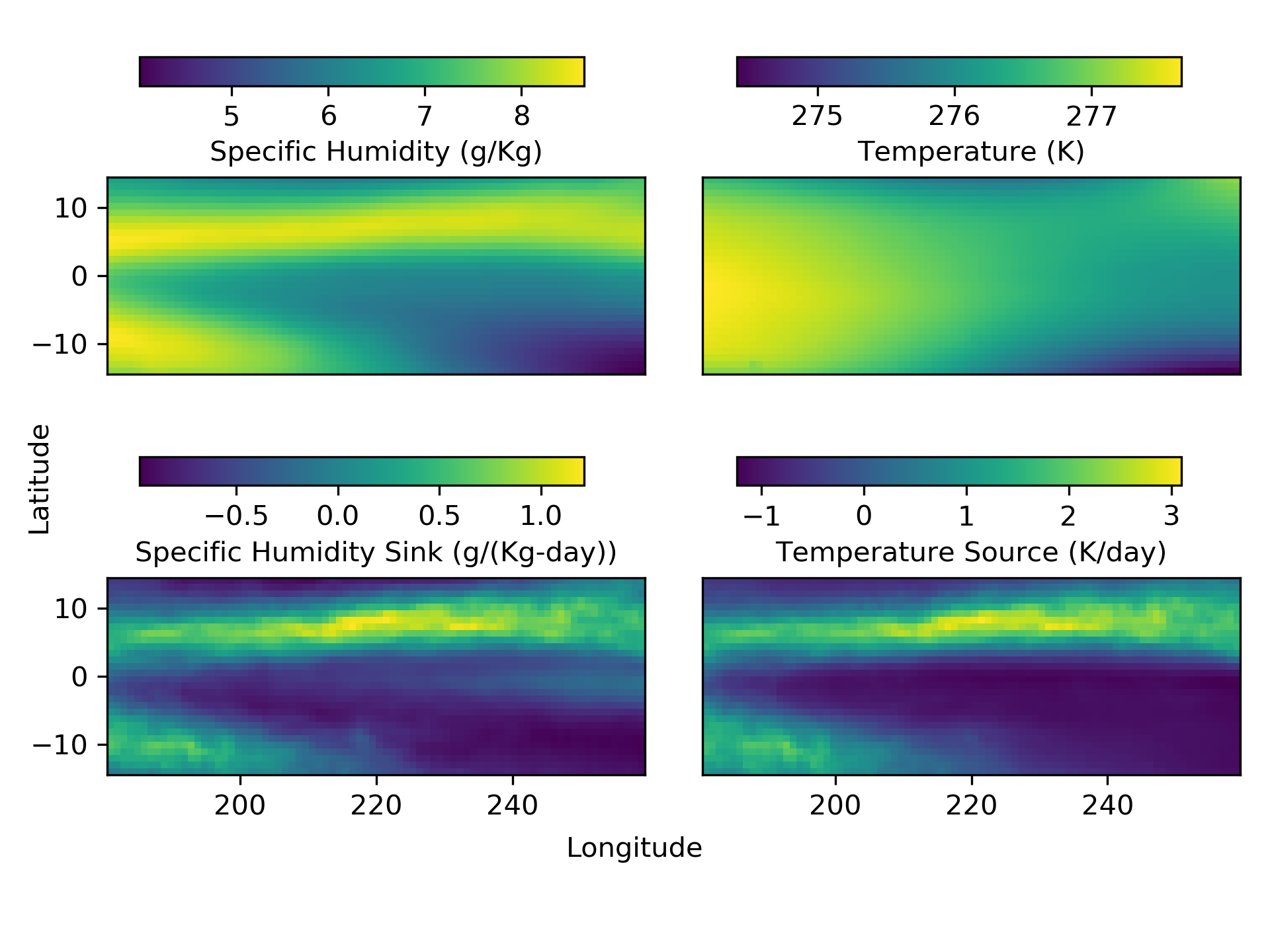}
  \makebox[-10pt][r]{% Similar to \llap
    \raisebox{2.5em}{%
      \includegraphics[trim=430 650 300 220, clip,
      width=.14\linewidth]{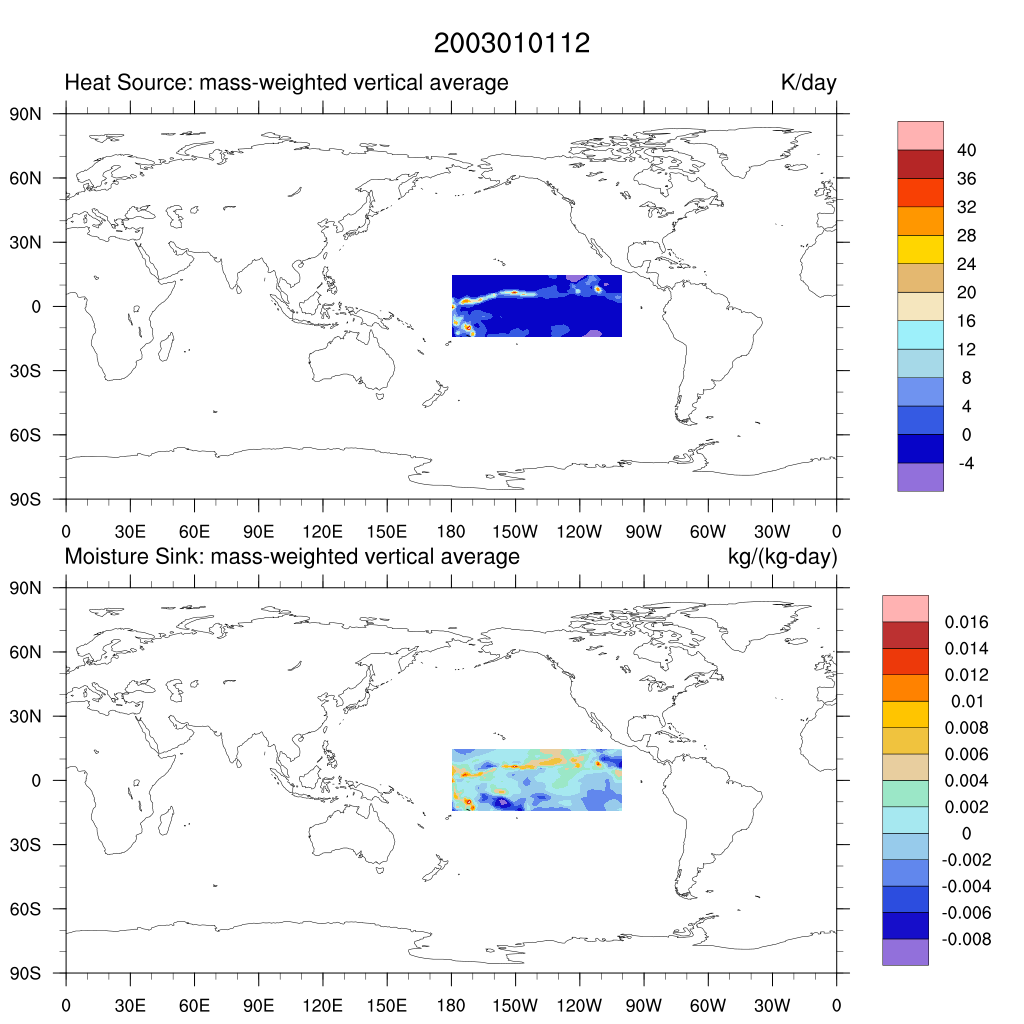}% Inserted image/inset
    }\hspace*{1em}%
  }%
  \end{center}
  {\caption{Preprocessing of MERRA2 data in the tropical Eastern
      Pacific. Top: Annual mean of column-averaged specific humidity
      (left) and temperature (right) in the Eastern Pacific. Bottom:
      Annual mean of apparent moisture sink (left) and apparent
      heating (right). To aid in geo-locating the domain considered
      and the ITCZ, the inset in the bottom-right panel shows the
      $q_1$ field for January 1$^{st}$, 2003 on a map}
    \label{fig1}}
\end{figure}

Whereas the timescales of convective updrafts range from minutes to
hours, the reanalyzed fields under NASA’s Modern-Era Retrospective
analysis for Research and Applications, version 2 (MERRA2) project are
made available at a sampling interval of 3 hours. Nevertheless, the
reanalyzed fields contain the averaged effects of convective updrafts
and other subgrid processes.  First, we use the $T$, $q$, and
($u, v, \omega$) fields to estimate $q_1$ and $q_2$ using Eqs.~1 and
2. On so doing, we find that there are numerous instances wherein the
diagnosed $q_1$ and $q_2$ fields are dominated by the vertical
advection of temperature and specific humidity respectively, and which
are themselves much larger than the horizontal advection terms and the
temporal tendency terms. Furthermore, in such cases, the vertical
distributions of $q_1$ and $q_2$ were found to be highly correlated to
the vertical distributions of vertical advection of temperature and
specific humidity respectively. We proceed to further investigate this
aspect of the diagnosed diabatic sources in order to establish that
they are not mere artifacts before proceeding with modeling them.

MERRA2 is a reanalysis product in the sense that it uses an
atmospheric general circulation model (AGCM), the Goddard Earth
Observing System version 5 (GEOS-5) and assimilates observations to
produce interpolated and regularly-gridded data
\citep{rienecker2011merra}.  While vertical motion is intimately
linked to the stability of the atmosphere given stratification in the
vertical, and is key to the formation of cumulus, vertical velocity is
usually not directly measured.  It is indeed the case that direct
observations of vertical motion are not part of the data that are
assimilated by the GOES-5 AGCM to produce MERRA2. As such, to examine
whether the high frequency (temporal), high-wavenumber (spatial)
component of the inferred vertical (pressure) velocity of MERRA2
(components in which the confidence tends to be lower than in the
slower, larger-scale components) leads to the correlation mentioned
above, we consider further spatiotemporal averaging of the MERRA2
fields from a native resolution of 0.5 $\times$ 0.625 degrees at 3
hours to a daily average at a coarsened resolution of 1 $\times$ 1
degree \citep[e.g.,][]{sun2015isolating}. The persistence of the
correlations, however, suggests that the correlations are likely real.

All fields are considered on MERRA2's native vertical grid (pressure
levels) and we confine attention to the troposphere, defined here for
convenience as pressure levels deeper than 190 mbar (hPa); there are 31 such
levels.  Figure~1 shows the annual and vertical average of specific
humidity (top-left), temperature (top-right), diagnosed specific
humidity sink (bottom-left), and temperature source
(bottom-right). The ITCZ is seen to be better defined in the diagnosed
(bottom) fields and its structure in $q_1$ and $q_2$ is also seen to
be highly correlated.

\section{Generative Adversarial Networks}
The governing equations would be closed if we are able to express the
three-dimensional fields $q_1$ and $q_2$ in terms of the resolved (3D)
variabless; given that they are not closed, we proceed to find such
closures using machine-learning. However, given the physics of the
problem described previously, the dominant variation of such
relationships is captured by the column-wise variation of $q_1$ and
$q_2$ in terms of the column-wise variation of $T$ and $q$.
Furthermore, as alluded to earlier, we should not expect unique
functional relationships between the predictands and predictors. As
such a probabilistic ML framework is called for. That is, instead of
learning maps
$$[T(p), q(p)] \mapsto [q_1(p), q_2(p)],$$ we are led to want to
learn the conditional probability
$$P\left([q_1(p), q_2(p)] \mid [T(p), q(p)]\right),$$ the probability
of $[q_1(p), q_2(p)]$ occuring given the occurence of conditions
$[T(p), q(p)]$. 

In this setting, a conditional generative model is required, since in
essence, a generative model learns a representation of the {\em
  unknown probability distribution} of the data.  Different classes of
generative models have been developed and they can range from
generative adversarial networks (GANs)
\citep{goodfellow2014generative} to variational autoencoders
\citep{kingma2013auto} to continuous normalizing flows
\citep{rezende2015variational} to diffusion maps
\citep{coifman2006diffusion} and others. However, we choose GANs since
they have been investigated more extensively than the others.  As a
class of generative models, GANs are trained using an adversarial
technique: while a discriminator continually learns to discriminate
between the generator’s synthetic output and true data, given a noise
source, the generator progressively seeks to fool the discriminator by
synthesizing yet more realistic samples. The (dual) learning
progresses until quasi-equilibrium is reached (see Fig. 2a for an
example of this in the current setup), wherein the discriminator's
feedback to the generator does not permit further large improvements
of the generator, but rather leads to stochastic variations that
explore the local basin of attraction. In this state, the training of
the generator may be considered as successful to the extent that the
discriminator has a hard time, both, discriminating between true samples and
fake samples created by the trained generator, and learning further to
discriminate them.

Initially, the training of GANs was problematic in that they commonly
suffered from instability and various modes of failure. One of the
approaches developed in response to the problem of instability in the
training of GANs was the Wasserstein GAN (WGAN)
\citep{arjovsky2017towards} The WGAN approach leverages the
Wasserstein distance \citep{kantorovitch1958translocation} on the
space of probability distributions to produce a value function that
has better theoretical properties, and requires the discriminator
(called the critic in the WGAN setting) to lie in the space of
1-Lipschitz functions. While the WGAN approach made significant
progress towards stable training of GANs and gained popularity,
subsequent accumulating evidence revealed that the WGAN approach could
sometimes generate only poor samples or, worse, even fail to converge.
In further response to this aspect of WGANs, a variation of the WGAN
was proposed in \cite{gulrajani2017improved}: while in the original
WGAN, a weight-clipping approach was used to ensure that the critic
lay in the space of 1-Lipschitz functions, the new approach used a
gradient penalty term in the critic's loss function to satisfy the
1-Lipschitz requirement. As such this approach was called the
Wasserstein GAN with gradient penalty (WGAN-GP).

\subsection{Architecture and Loss Functions}
In the conditional version of the WGAN-GP we use, besides the input
and output layers, the generator and critic networks each consist of
three fully connected feed-forward hidden layers (multi-layer perceptron
architecture).  The hidden layers have 512 neurons each with 10\%
dropout, and use rectified linear unit (ReLU) activation functions while the
input and output layers are linear. The noise sample was concatenated to the
condition vector for input to the generator. And, the condition vector
was concatenated to the real or generated sample to form the input to
the critic network. 

The dataset consisted of 864,000 datapoints (30 latitude bins, 80
longitude bins, and 360 daily averages (1/4-12/29)) with 62 conditions
representing temperature and humidity at 31 vertical pressure levels.
The target variables were $q_1$ and $q_2$ at the same 31 vertical
levels. To account for the typical dominant nature of the seasonal
component of climate variability, the dataset was split into quarters,
the first 70 days of each quarter were used for training and the
subsequent 20 days of each quarter constituted the test set. Training
data was scaled to a mean of zero and unit standard deviation
(standardized), and the same, learned, transformation was applied to
the test data.  Data was inverse transformed prior to plotting and
metric calculations.  Training data was split into 100 mini-batches per
epoch.

The generator and critic loss functions were as proposed in
\cite{gulrajani2017improved} with gradient penalty added to the critic
loss. Training was performed with an Adam optimizer with momentum
parameters ($\beta$s) set to 0.5 and 0.9. The coefficient of gradient
penalty was set at 0.1. We note that when using the same generator and
critic architectures, but with other data \citep{hossain2021high}, we
have found robust behavior of the predictions over a wide range of
values for the coefficient of gradient penalty (0.01 to 10).  A
learning rate of 1e-3 was used. The generator was only updated every 5th
step in order to allow the critic to learn the Wasserstein distance
function better before updating the generator.  Once training was
complete, the test conditions along with noise samples were fed
to the generator to produce results that we discuss in the next
section.

Variations of hyperparameter considered consisted of number of neurons
in each layer, number of layers, activation function, coefficient of
gradient penalty in discriminator loss, dropout and batch
normalization layers in generator, and the dimension of the noise
vector input to the generator. No sensitive dependencies were
found. Furthermore, a validation set was not used for the reason that
the training of the cGAN was stable and no special stopping
criterion was used. For example, the averaged
generator and discriminator loss functions were monotonic and could be
continued for tens of thousands of epochs (see Fig. 2a for an
example).

\begin{figure}
    \centering
    \begin{subfigure}[b]{0.5\textwidth}
            \centering
            \includegraphics[width=\textwidth]{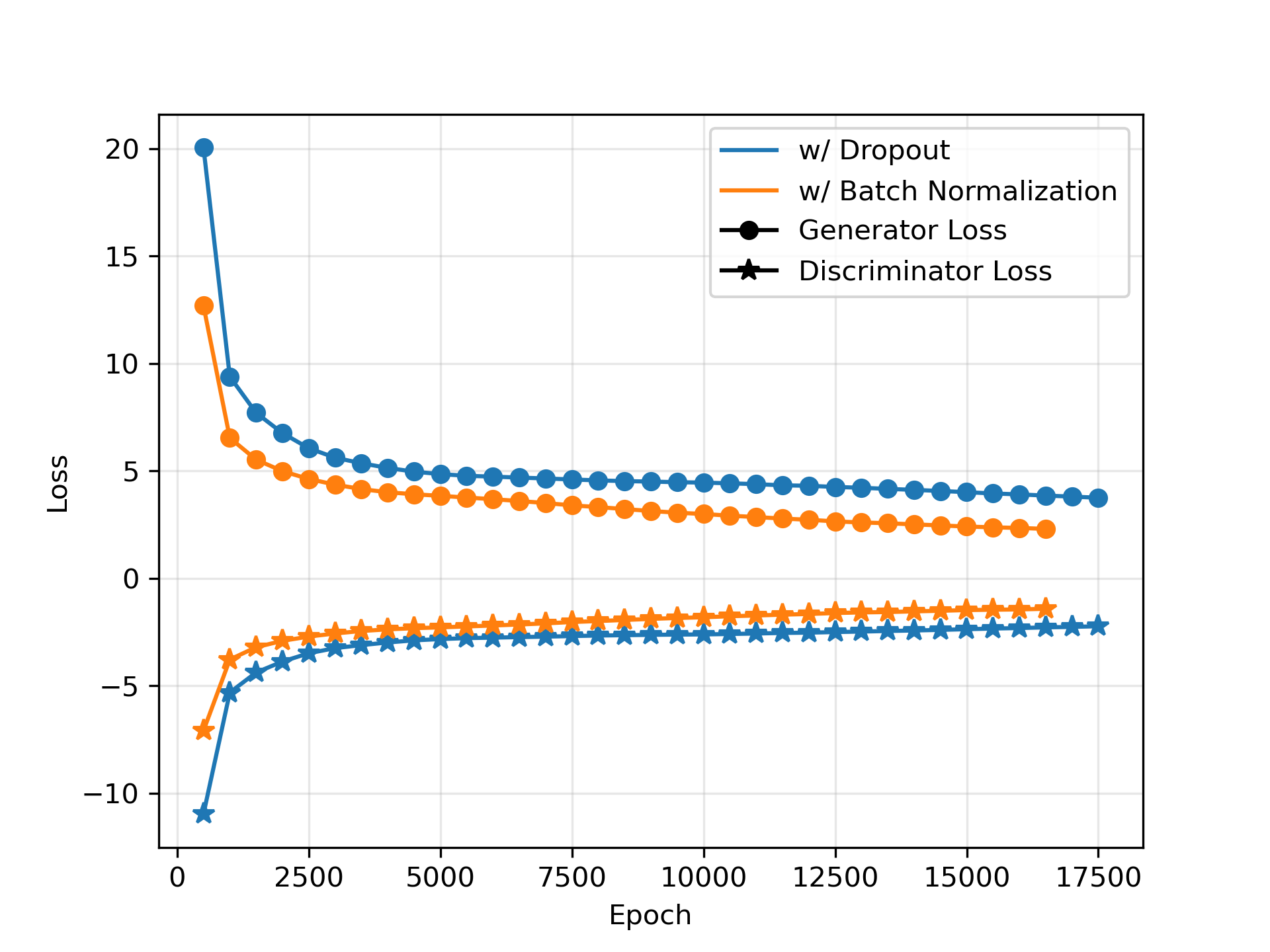}
            \caption{}
    \end{subfigure}%
    \begin{subfigure}[b]{0.5\textwidth}
            \centering
            \includegraphics[width=\textwidth]{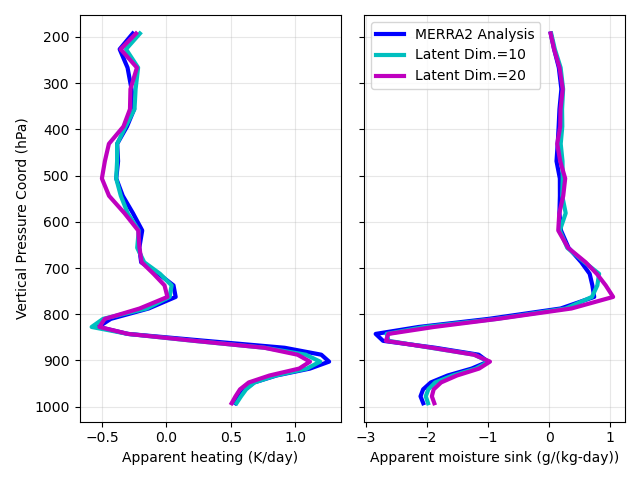}
            \caption{}
          \end{subfigure}
    \caption{(a): Training loss with dropout (blue) and with batch
      normalization as a function of training epoch. Both generator
      and discriminator losses are shown 
      and the training is seen to continue stably over large numbers
      of epochs. (b): Comparison of vertical distribution of cGAN predictions
      against reference MERRA2 analysis. Averages over test set
      of apparent heating (left) and apparent moisture sink (right)
      are compared. Teal (Magenta) values are for the case when the
      dimension of the noise vector is ten (twenty)}
\label{fig2}
\end{figure}

\begin{figure}
    \centering
    \begin{subfigure}[b]{0.5\textwidth}
            \centering
            \includegraphics[width=\textwidth]{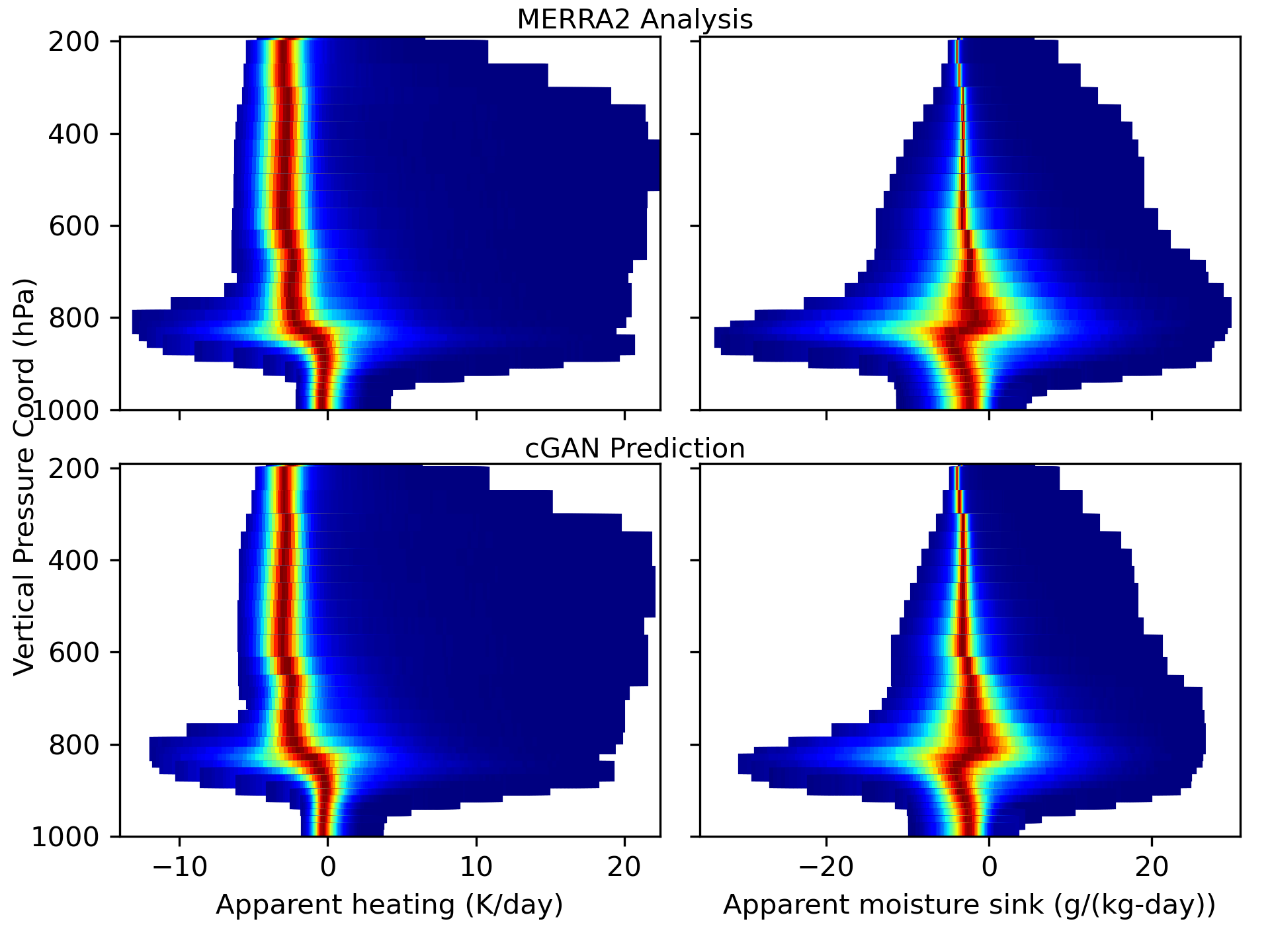}
            \caption{}
    \end{subfigure}%
    \begin{subfigure}[b]{0.5\textwidth}
            \centering
            \includegraphics[width=\textwidth]{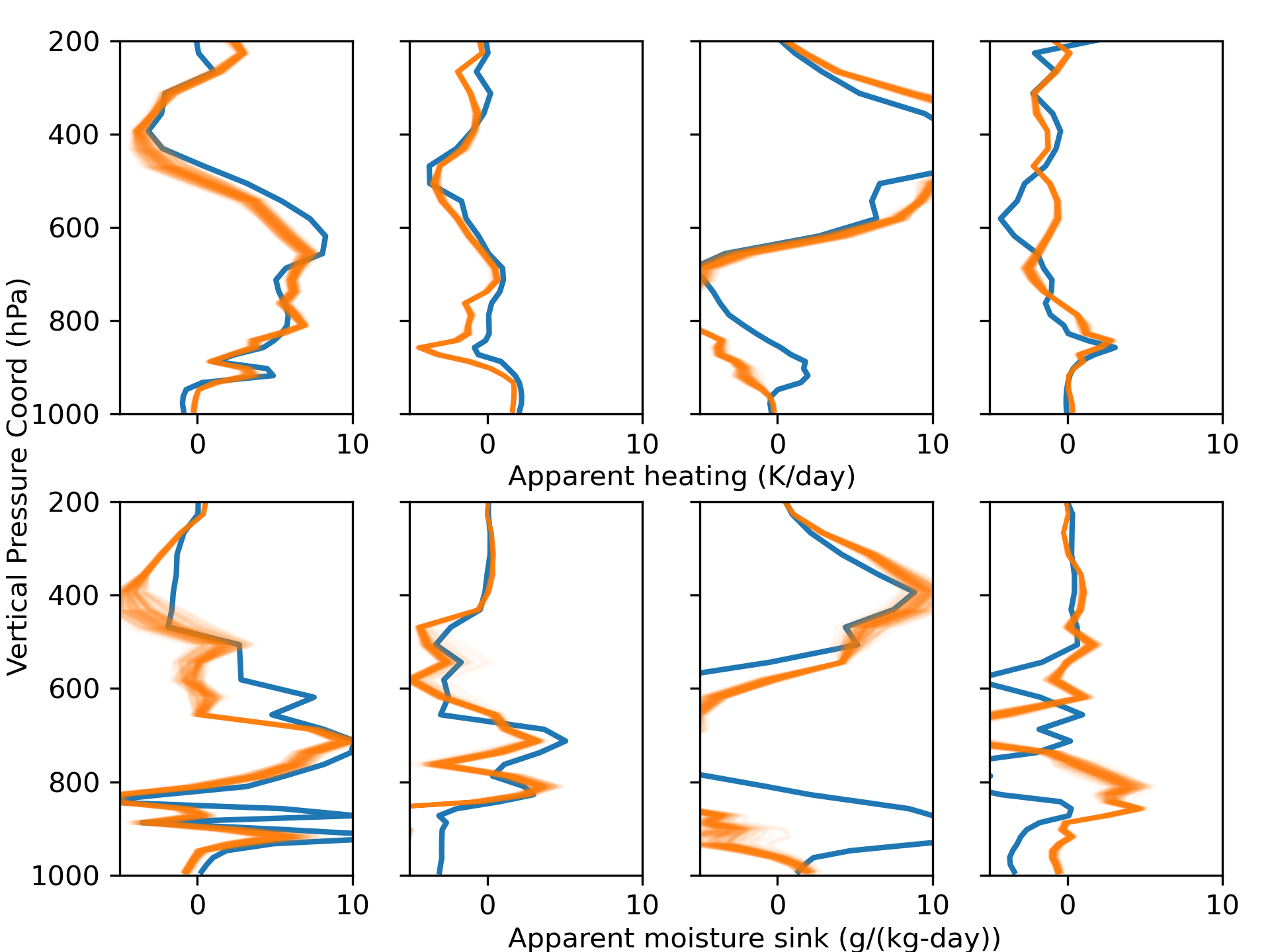}
            \caption{}
    \end{subfigure}
    \caption{(a): The
      probability distribution functions (pdf; red: high probability;
      blue: low probability) over the test set are computed at each
      pressure level and compared against the MERRA2 pdfs. The
      variable depth of the MERRA2 vertical layers is apparent in
      these figures. Apparent heating is shown in the left column and
      apparent moisture sink is shown in the right column. Reference
      MERRA2 analysis is shown in the top row and cGAN predictions
      (noise vector dimension of ten) are shown in the bottom row.
      (b): A stochastic ensemble of cGAN predictions are shown for four
  random instances in the test set. The top row is for heating whereas
the bottom row is for moisture sink. The reference MERRA2 profiles are shown in
blue and the predictions in orange.}
\label{fig3}
\end{figure}

\section{Results}
Given the physics of the problem, and as described earlier, we focus
on the vertical distributions of apparent heating and moisture sink,
conditioned on the vertical distributions of temperature and specific
humidity. Figure~2b shows comparisons of the vertical distribution of
cGAN predictions against reference MERRA2 analysis. Here, averages
over the test set are compared, with apparent heating shown in the
left panel and the apparent moisture sink shown in the right panel.
Two sets of cGAN predictions are shown with the only change being the
dimension of the noise vector (cyan: 10, magenta: 20).  The overall
structure of the vertical distribution of both heating and moisture
sink are reasonably well captured by the cGAN predictions with no
indication of bias.

In more detail, in Fig. 3a, we consider the probability distributions
over the test set at each pressure level with red indicating high
probability and blue indicating low probability. Apparent heating is
shown in the left column and apparent moisture sink in the right
column. Reference MERRA2 analysis is shown in the top row and cGAN
predictions (noise vector dimension ten) in the bottom row. The pdfs
of the cGAN predictions are seen to compare well with the refernce
MERRA2 analysis. In particular, the predictions capture the skewed
nature of the distributions well. However, the regions of high
probability (width of red regions) in the pdfs of the cGAN predictions
of heating, are seen to be slightly sharper than in MERRA2
analysis. (With batch normalization, while the width of the mode
region is as in MERRA2, a different minor discrepancy is seen: there
are sharper jumps in the vertical.)  For reference, at a yet more
detailed level, we shows stochastic ensembles of predictions at four
randomly chosen test conditions in Fig.~3b. The reference MERRA2
profiles are shown in blue and the predictions in orange. . Note that
while the cGAN predictions sometimes appear as a thick orange line,
they are actually an ensemble of thin lines. The stochastic ensemble
was obtained by holding the test condition fixed while repeatedly
sampling the noise vector. More often, the cGAN predictions are seen
to better capture the slower variations.

\section{Conclusion}
The tropical Eastern Pacific is an important region for global climate
and contains regions of intense tropical convection---segments of the
ITCZ. Adopting a framework of apparent heating and apparent moisture
sink to approximately separate the effects of dynamics and
column-physics, we diagnosed the ``physics'' sources using the MERRA2
renalysis product in a preprocessing step. We
performed the analysis at MERRA2's native spatiotemporal resolution,
and then at a significantly coarser resolution.
In both cases, we found that the heating and moisture sink were highly correlated
with vertical advection of temperature and humidity ($\sim$ 0.9, and $-0.7$) respectively.
% We reasoned that confidence in the high-frequency
% high-wavenumber component of the MERRA2 estimate of
% vertical (pressure) velocity is likely not high since direct
% measurements of vertical velocity are not assimilated into the GOES-5
% AGCM used to produe MERRA2. Indeed, upon conducting the analysis after
% MERRA2 data was subject to further spatiotemporal averaging, no such
% anomalous behavior was seen.

Thereafter, we designed a probabilistic machine learning technique to
learn the distribution of the heating and moisture sink profiles
conditioned on vertical profiles of temperature and specific
humidity. This was based on a Wasserstein Generative Adversarial
Network with gradient penalty (WGAN-GP). After successfully training
the cGAN, we were able to demonstrate the viability of the methodology
to learn unified stochastic parameterizations of column physics by
comparing predictions using test conditions against reference MERRA2
analysis profiles. We expect this work to add to the growing body of literature
\citep[e.g.,][and others]{krasnopolsky2010accurate,
  krasnopolsky2013using, rasp2018deep,
  o2018using,brenowitz2018prognostic} that seeks to develop
machine-learned column-physics parameterizations towards alleviating
both, shortcomings of human-designed physics parameterizations, and
the computational demand of the “physics” step in climate models.

We conclude by mentioning a few issues that need further
investigation. Given the diversity of behavior (e.g., ITCZ
vs. non-ITCZ regions), it remains to be seen if adding velocity fields
as predictors will improve performance. Training over ITCZ regions and
testing elsewhere and other such combinations are likely to render
further insight into the parameterizations. The stochastic ensemble our
method produces, e.g., like in Fig.~3b, can be readily subject to
reliability and other analyses using tools developed in the
context of probabilistic weather forecasts \cite[e.g.,
see][and references therein]{nadiga2013ensemble, luo22}. We anticipate that such
analyses will help us better understand the role played by the
dimension of the latent space used in generative modeling.  Finally,
stability of the machine learned parameterization in a GCM setting is
an important issue. For example, \cite{brenowitz2018prognostic} find
this to be a problem when they learn {\em deterministic}
parameterizations from CRM-based data. They circumvent the problem by
changing the problem formulation to include time integration of the
temperature (in their case static stability) and moisture equations
and changing the loss function to penalize deviations of the
predictions of $T$ and $q$ at multiple times in a cumulative sense. It
remains to be seen if the {\em stochastic} parameterizations learned
using a probabilistic ML methodology, as we do presently, can provide
an alternative way of addressing the stability issue.

\begin{Backmatter}
\paragraph{Funding Statement}
This research was supported under U.S. Department of Energy (DOE), Office of
Science's Scientific Discovery through Advanced Computation (SciDAC4)
program under project ``Non-Hydrostatic Dynamics with Multi-Moment
Characteristic Discontinuous Galerkin Methods (NH-MMCDG)''.

\paragraph{Competing Interests}
None

% A statement about how to access data, code and other materials
% allowing users to understand, verify and replicate findings ---
% e.g. Replication data and code can be found in Harvard Dataverse:
% \verb+\url{https://doi.org/link}+.
\paragraph{Data Availability Statement}
  MERRA2 data used in this work is available and can be downloaded at https://disc.gsfc.nasa.gov/datasets?project=MERRA-2

\paragraph{Author Contributions}
Conceptualization:
BTN, Methodology: BTN, CN. Data curation: BTN, XS. Data visualisation:
BTN Writing original draft: BTN. All authors approved the final
submitted draft. 
% \paragraph{Supplementary Material}
% State whether any supplementary material intended for publication
% has been provided with the submission.
\bibliography{refs}
\bibliographystyle{apalike}

\end{Backmatter}

\end{document}